\documentclass[12pt,preprint]{aastex}

\usepackage{amsmath}
\usepackage{graphicx}
\usepackage{epsfig}
\usepackage{bm}
\usepackage{color}

\newcount\doepsf
\newcommand{\be}{\begin{equation}}
\newcommand{\ee}{\end{equation}}
\newcommand{\bea}{\begin{eqnarray}}
\newcommand{\eea}{\end{eqnarray}}
\newcommand{\bean}{\begin{eqnarray*}}
\newcommand{\eean}{\end{eqnarray*}}

\newcommand{\aanda}{{A\&A}}

\newcommand{\pp}{{Phys. Plasmas}}

\begin{document}

\title{IMPACT OF TEMPERATURE DEPENDENT RESISTIVITY AND THERMAL
  CONDUCTION ON PLASMOID INSTABILITIES IN CURRENT SHEETS IN
  SOLAR CORONA}

 
 

\author{Lei Ni\altaffilmark{1,2}, Ilia I. Roussev\altaffilmark{3,1},
Jun Lin\altaffilmark{1}, and Udo Ziegler \altaffilmark{4}}

\altaffiltext{1}{Yunnan Astronomical Observatory, CAS, P.0.
Box 110, Kunming 650011, Yunnan, China; leini@ynao.ac.cn}

\altaffiltext{2}{Key Laboratory of Solar Activity, National
Astronomical Observatories, Chinese Academy of Sciences,
Beijing 100012, China; leini@ynao.ac.cn}

\altaffiltext{3}{Institute for Astronomy, University of Hawai'i,
2680 Woodlawn Dr, Honolulu, Hawaii 96822, USA;
iroussev@ifa.hawaii.edu}

\altaffiltext{4}{Leibniz-Institut f\"ur Astrophysik Potsdam,
DE 14482 Potsdam, Germany; uziegler@aip.de}

\begin{abstract}

In this paper we investigate, by means of two-dimensional magnetohydrodynamic simulations, the impact of temperature-dependent resistivity and thermal conduction on the development of plasmoid instabilities in reconnecting current sheets in the solar corona.  We find that the plasma temperature in the current sheet region increases with time and it becomes greater than that in the inflow region.  As secondary magnetic islands appear, the highest temperature is not always found at the reconnection $X$-points, but also inside the secondary islands.  One of the effects of anisotropic thermal conduction is to decrease the temperature of the reconnecting $X-$points and transfer the heat into the $O-$points, the plasmoids, where it gets trapped.  In the cases with temperature-dependent magnetic diffusivity, $\eta \sim T^{-3/2}$, the decrease in plasma temperature at the $X-$points leads to: (i) increase in the magnetic diffusivity until the characteristic time for magnetic diffusion becomes comparable to that of thermal conduction; (ii) increase in the reconnection rate; and, (iii) more efficient conversion of magnetic energy into thermal energy and kinetic energy of bulk motions.  These results provide further explanation of the rapid release of magnetic energy into heat and kinetic energy seen during flares and coronal
mass ejections.  In this work, we demonstrate that the consideration of anisotropic thermal conduction and Spitzer-type, temperature-dependent magnetic diffusivity, as in the real solar corona, are crucially important for explaining the occurrence of fast reconnection during solar eruptions.

\end{abstract}


\section{INTRODUCTION}
\label{sec:introduction}

Recent advanced solar observations suggest that plasmoids are
ejected from reconnection sites in the solar corona both away and
toward the Sun during Coronal Mass Ejections (CMEs), or solar
eruptions \cite[]{Savage10,Nishizuka10,Milligan10, Lin05}.  Stemming
from these observations, we can assume that the CME's current sheet
is not a single layer of enhanced current density, but it contains
many fine structures within, including multiple reconnection X-points
\cite[]{Lin08}.  Since the majority of space plasma systems are
collisionless, it is important to study the reconnection dynamics via
the kinetic approach.  The complexity of the underlying physics limits
kinetic models and simulations of reconnection in these space
environments to relatively small regions (size of ion-inertia length),
which are currently under resolved with the existing observational
facilities.  The collisional theory, however, can still be used in
numerous space plasma physics circumstances in order to calculate
the reconnection rate, as well as the rate at which the magnetic
energy is dissipated.  Magnetohydrodynamic (MHD) reconnection
solutions exist where the rate of reconnection is largely independent
of the magnitude of the electric resistivity \cite[]{Priest00}.  In
physical circumstances high Lundquist numbers ($S \gtrsim 10^4$),
existing numerical simulations have already demonstrated that a single
reconnecting current sheet can break up into multiple interacting
reconnection sites even on the MHD scale \cite[]{Biskamp86,
Bhattacharjee09,Huang10,Shen11,Barta11,Mei12}.  It was also found
that, should the aspect ratio of the secondary current sheet exceed
some critical value ($\gtrsim 60$), then higher orders of magnetic
islands (or O-points) and thinner current sheets begin to develop
\cite[]{Ni10,Ni12}.  As a result, the local value of current density
at the X-points and the global reconnection rate can be increased
significantly during the reconnection process involving plasmoid
instabilities.  This yields fast reconnection dynamics in the physical
environment of the solar corona, as required to explain solar
observations of flares and CMEs.     

In the majority of existing MHD numerical studies of plasmoid instabilities,
for simplicity, the magnetic diffusivity coefficient is assumed to be
either uniform, or a function of position.  For collisional space
plasma on the MHD scale, it is well established, however, that the
magnetic diffusivity varies with the plasma temperature approximately
as: $\eta \sim T^{3/2}$ \cite[]{Spitzer62,Schmidt66}.  Since the
plasma temperature in reconnecting current sheets is generally not
uniform (varies with time and location), studying the reconnection
process with a realistic, temperature-dependent magnetic diffusion
is very important.  Furthermore, the effects of thermal conduction on
the reconnection dynamics in current sheets have also been ignored in
the existing numerical 2-D (and 3-D) MHD models.  Some previous
studies, however, indicate that this term could be very important
\cite[]{Takaaki97,Chen99,Botha11}, should the temperature and its
spatial gradient be high enough in the underlying physical
environment, such as that of the solar corona. 

\textcolor{blue}{On the choice of resistivity model in our simulations, note that there are numerous MHD models in the literature that adopt some (either ad-hoc, or physics-based) form of anomalous resistivity to obtain fast reconnection. For example, the MHD works of Roussev et al.\cite[]{Roussev02} and B\'arta et al.\cite[]{Barta11} adopt physics-based models of anomalous resistivity to investigate the dynamics of fast reconnection in current sheets. In these studies, the anomalous resistivity is triggered by the drift velocity or electric currents exceeding some critical values.  Some other studies \cite[]{Buchner06, Nishikawa96} have utilized particle-in-cell (PIC) codes to explore the nature of anomalous resistivity in reconnecting current sheets. What has been found so far is that the exact form(s) of anomalous resistivity used in the MHD models are not directly deducible from the kinetic theory and simulations of real physical systems, and therefore some simplifying assumptions are still required for the model of anomalous resistivity.  For these reasons, we have refrained from using any model of anomalous resistivity in our high-$S$-number studies, and we demonstrate here that fast reconnection in the solar corona can be achieved even with the
classical Spitzer-type resistivity.} 

In this paper, we investigate numerically the physical effects of
temperature-dependent magnetic diffusivity and anisotropic thermal
conduction on the evolution of plasmoid instabilities in reconnecting
current sheets in the low solar corona.  We analyze in great detail
the spatial and temporal evolution of the current density, magnetic
flux, reconnection rate, and temperature structure of reconnecting
current sheet during the development of plasmoid instability process.
The energy conversion process is also studied in great length and
compared in the cases with and without thermal conduction.  The
organization of the paper is as follows.  In Section 2, we present the
resistive 2-D MHD equations utilized in this work, as well as the
chosen initial and boundary conditions for the numerical experiments.
Our scientific findings are presented and discussed in Section 3.
Lastly, in Section 4, we summarize the results of our work and we
outline future plans for research relevant to the subject.


\section{FRAMEWORK OF NUMERICAL MODELS}
\label{sec:models}
The MHD equations describing the physical evolution of the low solar
corona, including the effects of magnetic diffusion and anisotropic
thermal conduction, are given by:

\begin{equation}
 \partial_t \rho = -\nabla \cdot (\rho \textbf{v}),
\end{equation}

\begin{equation}
 \partial_t e = -\nabla \cdot [(e+p+\frac {1}{2 }\vert \textbf{B}
 \vert^2)\textbf{v}-
(\textbf{v} \cdot \textbf{B})\textbf{B}] + \nabla \cdot [\eta
\textbf{B} \times (\nabla \times \textbf{B})-\textbf{F}_{cond}],
\end{equation}

\begin{equation}
  \partial_t (\rho \textbf{v}) = -\nabla \cdot [\rho \textbf{v}
  \textbf{v}
+(p+\frac {1}{2} \vert \textbf{B} \vert^2)I - \textbf{B} \textbf{B} ],
\end{equation}

\begin{equation}
 \partial_t \textbf{B} = \nabla \times (\textbf{v} \times \textbf{B}-
\eta\nabla \times \textbf {B}),
\end{equation}
 
 \begin{equation}
   e = p/(\Gamma_0-1)+\rho \textbf{v}^2/2+\textbf{B}^2/2,
 \end{equation}
   
 \begin{equation}
     p =   \rho T,
 \end{equation}
 
 \begin{equation}
   \textbf{F}_{cond} = -\kappa_{\parallel}(\nabla {T} \cdot \hat{\textbf{B}}){\hat{\textbf{B}}} -    
   \kappa_{\perp}(\nabla {T} - (\nabla {T} \cdot \hat{\textbf{B}})\hat{\textbf{B}}).
 \end{equation}
 
The above equations are solved in 2-D space using the NIRVANA code
(version 3.5) \cite[]{Ziegler08}, and all the variables therein are
dimensionless.  The simulation domain ranges from $0$ to $1$ $(l_x=1)$
in the $x$-direction, and from $0$ to $4$ ($l_y=4$) in the
$y$-direction.  Here, $\rho$ is the plasma mass density, $e$ is the
total energy density, $\textbf{v}$ is the flow velocity, $\textbf{B}$
is the magnetic field, $\hat{\textbf{B}}={\textbf{B}}/{\vert \textbf{B} \vert}$
is the unit vector in the direction of the magnetic field, $T$ is the
temperature, $p$ is the plasma thermal pressure, and $\eta$ is the
normalized magnetic diffusivity.  Note here that $\eta$ can be chosen
to be either uniform, or temperature-dependent in our models.  The
Lundquist number is defined by: $S=l_y v_{Ad}/\eta$, where $v_{Ad}$ is
the initial normalized asymptotic Alfv\'en speed at the upstream
boundary, which is set to 1.0 in all our models.  The plasma is
considered to be a fully ionized hydrogen gas, and the kinetic
temperatures of ions and electrons are assumed to be equal.  The
ratio of specific heats, $\Gamma_0$, is set to $5/3$ (ideal gas).  The
parallel, $\kappa_{\parallel}$, and perpendicular, $\kappa_{\perp}$,
thermal conductivity coefficients, in normalized form, are given by
the Spitzer theory \cite[]{Spitzer62}:

\begin{equation}
    \kappa_{\parallel} = c_1\cdot \kappa_{SP},
 \end{equation}
  
 \begin{equation}
   \kappa_{\perp} = c_2 \cdot 8.04 \times 10^{-33} (\frac
   {\ln{\Lambda}}{m_u})^2   \frac{{\rho}^2}{T^3 B^2} \kappa_{SP}.
 \end{equation}
 
Here, $\kappa_{SP} = 1.84 \times 10^{-10}/({\ln{\Lambda}} T^{5/2})$ is
the Spitzer's thermal conductivity coefficient,  $\ln{\Lambda}$ is the
Coulomb logarithm set to $30$, $m_u = 1.66057 \times 10^{-27}kg$
is the atomic mass unit,  $c_1 = \mu_0T_N^{7/2} /(B_N^2L_Nv_A)$, and
$c_2 = \mu_0 T_N^{1/2}\rho_N^2/(B_N^4 L_N v_A)$.  Also, $T_N$, $B_N$,
$L_N$ and $\rho_N$ are the normalization units for temperature,
magnetic field, length, and mass density, respectively, of the system.
The Alfv\'en speed is defined by $v_{AN} = B_N/\sqrt{\mu_0 \rho_N}$,
where $\mu_0 = 4\pi \times 10^{-7}$.  Normally, the choice of
normalization units is unimportant for resistive MHD, and the
equations can be solved in normalized form for any values for
$T_N$, $B_N$, $L_N$ and $\rho_N$.  This is not true, however, for
simulations that include the thermal conduction.  From the expression
of the heat conduction term, one can see that the physical values of
the normalization units, along with the temperature gradient,
determine the importance of this physical process.  The importance
of the cross-field thermal conduction coefficient, $\kappa_{\perp}$,
is restricted to the strong magnetic field case.  In the limit of
vanishing field strength, the heat conduction becomes isotropic and
$\kappa_{\perp}$ = $\kappa_{\parallel}$.  In our model, this is
accounted for by modifying $\kappa_{\perp}$ to be: $\kappa_{\perp} =
min( \kappa_{\perp}, \kappa_{\parallel} )$.  As a result, the
cross-field heat conduction coefficient cannot be greater than the
parallel one.  

In our models, we consider a Harris current sheet as the initial
condition for the magnetic field: 
 \begin{equation}
    B_{y0} = b_0 \tanh(\frac{x-0.5}{\lambda}), \quad B_{x0} = 0.
  \end{equation}   
Here $\lambda$ is the width of the current sheet set to $0.05$,
and $b_0 = 1$.  Note that the Harris current sheet should be thin
enough to enable tearing instabilities to develop according to the
criteria : $\frac{2}{\lambda} (\frac{1}{k\lambda}-k\lambda) > 0$,
where $k=2\pi/l_y$ is the wave number of the initial perturbations.
The initial velocity is set to zero in all simulations.  From Eq.~3,
the plasma pressure must satisfy the initial equilibrium condition,
which reads:
 \begin{equation}
    \nabla \cdot (p_0 \textbf{I}) = -\nabla \cdot [\frac{1}{2}\vert
    \textbf{B}_0 \vert^2   \textbf{I}-    
    \textbf{B}_0\textbf{B}_0].
 \end{equation}
 Since $\textbf{B}_0=B_{y0} \hat{\textbf{y}}$, where $\hat{\textbf{y}}$
is the unit vector in the $y$-direction,  the initial equilibrium plasma
pressure is calculated as:
   \begin{equation}
    p_0 = -\frac{1}{2}B_{y0}^2 + C_0, 
  \end{equation}  
where  $C_0$ is a constant.  From Eq.~10, we know that $B_{y0} = 1$
at the $x$-boundary.  Since the kinetic gas pressure is related to the
magnetic pressure by $\beta = \frac{p}{B^2/2} $, we obtain $C_0 =
\frac{\beta_0 +1}{2}$, where $\beta_0$ is the initial plasma $\beta$
at the $x$-boundary.  Thus:
  \begin{equation}
     p_0 = \frac{1+\beta_0-B_{y0}^2}{2}.
  \end{equation}
The initial equilibrium value of the total energy is:
  \begin{equation}
     e_0 = p_0/(\Gamma_0 -1)+B_{y0}^2/2.
  \end{equation}
 The initial temperature is assumed to be uniform in the entire
simulation domain.  From the ideal gas law $T=p/\rho$,  the
initial equilibrium mass density and temperature can be derived as:
 \begin{equation}
    \rho_{0} = p_{0}/T_{0}=\frac{1+\beta_0-B_{y0}^2}{\beta_0}, \quad
    T_{0}=\frac{\beta_0}{2}.
 \end{equation}
 
In order to trigger plasmoid instabilities in the current sheet, we
impose small initial perturbations for the magnetic field of the kind:
 \begin{equation}
     b_{x1} = -\epsilon \cdot 0.5 \sin(\pi x/l_x)\cos(2\pi y/l_y),
 \end{equation}
 \begin{equation}
    b_{y1} = \epsilon \cdot \cos(\pi x/l_x)\sin(2\pi y/l_y).
 \end{equation}
In our simulations, we used a constant value of $\epsilon = 0.05$.
The introduced perturbation has a half-period in the $x$-direction
and a full period in the $y$-direction.  This type of perturbation
yields the development of tearing-mode instabilities in the current
sheet and it produces a large primary magnetic island.  Eventually,
a much thinner Sweet-Parker-type current sheet develops, and
secondary magnetic islands appear if the Lundquist number is
sufficiently large ($\gtrsim 10^4$).  In all our simulations we
impose periodic boundary conditions in the $y$-direction and
Neumann boundary conditions in the the $x$-direction.  
 
We have simulated five different physical scenarios (models M0-M4
hereafter), which are discussed in this paper.  In models M0 and M1,
the magnetic diffusivity is considered to be uniform everywhere.  In
Models M2-M4, the Lundquist number scales with the temperature as
$S \sim T^{3/2}(x,y,t)$.  We choose $S=\frac{4}{6} \times 10^7 \times
T^{3/2}$ in order to make the Lundquist number sufficiently high to
yield the occurrence secondary plasmoid instabilities in the current
sheet.  The heat conducting term is included in models M1, M3 and
M4, and the choice of the normalization units affects the significance
of thermal conduction in the these cases. Table 1 summarizes the five
different models.  Note that the initial plasma $\beta_0$ is set to
0.2 in all the models.  The normalization unit for temperature is set
to $T_N=10^7$~K, and the initial temperature in the dimensional space
is $T_I = \frac{\beta_0T_N}{2}=10^6$~K for cases M1, M3 and M4, which
is similar to the temperature in the solar corona.  In the real solar
corona, the temperature could be higher than $10^6$~K, especially
within current sheets.  The magnetic field is of the order of $0.01$~T,
and the mass density is around 1-2 orders of magnitude smaller than
$9.576 \times 10^{-10} \, kg/m^3$,  so that the value of $c_1 = \mu_0
T_N^{7/2}/(B_N^2L_Nv_A)=\mu_0^{3/2}T_N^{7/2}\rho_N^{1/2}/(B_N^3L_N)$
in the real solar corona could be close to, or even greater than the
value of $c_1$ calculated for all the models.  For the choice of
normalization units in cases M1 and M3, we find that the magnitude
of the cross-field thermal conduction coefficient is around $10^8$
times smaller than the parallel one at the boundaries $x=0$ and
$x=1$.  The thermal conduction, however, is nearly isotropic initially
in the middle of the current sheet, because the magnetic field is weak
there. 
 
We perform the simulations on a base-level Cartesian grid of $80
\times 320$.  The highest refinement level in our simulations is 10,
which corresponds to a grid resolution of $\delta x = 1/81920$.  In
order to ensure that this resolution is sufficient, we have carried out
convergence studies starting with twice lower resolution.  In
specific, we have tested the case M3 by setting the highest refinement
level equal to 9, which corresponds to a grid resolution of $\delta_x
= 1/40960$.  We find that the reconnection rate is very similar in
both the high and the low resolution run. \textcolor{blue}{Hence,
the grid resolution in our simulations is sufficiently high to
suppress the effects of the numerical resistivity.}

\section{NUMERICAL RESULTS AND DISCUSSIONS}
\label{sec:results} 
In this paper, the time-dependent reconnection rate is defined as
$\gamma(t) = \partial (\psi_X(t) - \psi_O(t))/\partial t$, where
$\psi_X$ and $\psi_O$ are the magnetic flux functions at the
main reconnection $X$-point (where the separatrices separating
the two open field line regions intersect) and the $O$-point.  Here,
the magnetic flux function is defined through the relations: $B_x =
-\partial \psi/\partial y$, $B_y=\partial \psi/\partial x$.  The
$O$-point is always inside the primary island, and the
corresponding $\psi_O$ is the minimum value of $\psi$ over the
whole simulation domain.  In the case when there are several
$X$-points, the one which has the maximum value of $\psi$
dictates the reconnection rate.  Calculated this way, the reconnection
rate is the global one over the entire reconnecting current sheet.
 
While analyzing the data, we find the following key observations.
First, the temperature increases with time at the center of the
current sheet, especially at the reconnection $X$-point in the
beginning.  Since the plasma is ejected away from the $X$-point
during the development of plasmoid instability process, an increasing
amount of hot plasma gets trapped inside the magnetic islands.
The location of maximum temperature is sometimes found not to
be at the reconnection $X$-point, but inside the secondary islands.
The temperature in the current sheet region, however, is always
higher than in the inflow region.  Once can see this characteristic
evolution of the temperature in Fig.~1 and Fig.~2.
 
For a temperature-dependent magnetic diffusivity ($\eta \sim
T^{-3/2}$), $\eta$ decreases with increasing temperature.  Fig.~3(a)
and Fig.~3(b) illustrate the distributions of the current density and
the magnetic flux at different time instants for the case of uniform
magnetic diffusivity (M0), and the case of temperature-dependent
magnetic diffusivity (M2), respectively.  Since the initial
temperature, $T_0$, is set to 0.1 in all the cases we have simulated
here, the initial value of magnetic diffusivity is the same in all cases.
As the plasmoid instability develops with time in the case M2, the
magnetic diffusivity inside the current sheet decreases with
increasing temperature.  This makes the initial thick Harris sheet
evolve into a thinner Sweet-Parker current sheet, when compared
with the M0 case (with uniform magnetic diffusivity).  There also
appear more secondary islands and thinner secondary current sheets
in the case M2 than in the case M0.  The maximum current density at
the $X$-point in the case M2 can increase to higher values during the
secondary instabilities.  At the same time, however, the reconnection
rate and the maximum temperature in the simulation domain are
somewhat smaller than in the case M0.  These key observations can
be seen clearly in \textcolor{blue}{Fig.~2,} Fig.~3(a), Fig.~3(b), Fig.~4(a) and Fig.~4(b). \textcolor{blue}{Fig.~2 also shows that the temperature distribution along the current sheet is relatively smoother for the case of uniform magnetic diffusivity (M0).}
According to the normalization units chosen for cases M1 and M3,
the ion-inertia length is calculated to be around $100$~m in these
models.  The narrowest width of the secondary current sheets can
reach around 0.001 in our simulations, which corresponds to $0.001
L_C = 10^4$~m in the real space.  This is much greater than the
ion-inertia length, meaning that our simulations are in the
collisional regime.  Therefore, the adopted form of
temperature-dependent magnetic diffusivity is well justified in our
models. 

In the following we discuss the physical effects of thermal conduction
on the development of the plasmoid instabilities.  The numerical
results for the reconnection rate, the current density distribution,
the temperature distribution, and the energy conversion are presented
and compared for the cases with and without thermal conduction.

In the case with uniform magnetic diffusivity (M1), the heat conduction
(see Table 1 for characteristic parameters) does not affect significantly
the evolution of the plasmoid instability.  The reconnection rate, the
distribution of the current density, and the magnetic field structure
at each time step for the case M1 are very similar to those for the
case M0.  The cases M2, M3, and M4 have the same form of magnetic
diffusivity, which evolves with temperature as $\eta \sim T^{-3/2}$.
The thermal conduction term, however, is switched on only in cases
M3 and M4 (and turned off for M2).  Also, the normalization units are
the same in the cases M1 and M3, unlike the case M4 where the
normalization units for magnetic field strength and mass density are
chosen smaller.  This makes the value of $c_1$ greater in the case M4
than in the cases M1 and M3, which is why the thermal conduction
effects more pronounced in the former.  We find that the spatial and
temporal evolution of the current density and the magnetic flux to
be almost identical for the cases M2-M4 prior to the occurrence of
secondary magnetic islands.  As the plasma temperature and its
gradient increase in the current sheet during secondary instability
processes, the heat conduction effects become more pronounced
in the case M4 than in the other cases. \textcolor{blue}{Note that
the time-step becomes very small after $t=22t_A$ in the case M4,
which is why the simulation was terminated at this time.}

As far as the time-dependent current-sheet structure and reconnection
rate are concerned, one can see In Fig.~3(b) and Fig.~3(c) that the
secondary current sheets are thinner in the case M3 than in the case
M2 at the same time instant.  We also find that in the former case the
maximum current density at the reconnection $X$-point increases up
to a value is twice greater than in the latter case.  Fig.~4(a) reveals that
the reconnection rate is also almost twice greater in the case M3 than
in the case M2 during the later stages of the secondary instability
process.  In the case of M4, the reconnection rate can increase to an
even higher value compared to the cases M2-M3.  As seen in Fig.~2(b),
when there are secondary plasmoids present in the current sheet, the
temperature distribution along the current sheet is not smooth
anymore, and there are several temperature peaks inside the secondary
plasmoids.  This makes the temperature gradients along the current
sheet become large enough to render the heat conduction important.
This leads to a further increase in the reconnection rate, as seen in
the case of M4.

In order to demonstrate the significance of thermal conduction, we
calculate the characteristic time-scale of heat conduction , $t_{TH}
\sim E_{th} / (\kappa_{\parallel} \partial^2T/ \partial x^2)$ along
the current sheet in the case M3 at $t=22.534t_A$.  Here, $E_{th}$
is the thermal energy density, $\kappa_{\parallel}$ is the field-aligned 
thermal conduction coefficient, and $ \partial^2T / \partial x^2 $ is
the second derivative of plasma temperature.  We find that $t_{TH}$
is significantly shorter ($\sim 0.1 t_A$) than the Alfv\'en crossing
time at the reconnection $X$-point locations, as well as the
$O$-point locations than elsewhere in the current sheet.  Note,
however, that at the locations of the $O$-points the temperature
gradient is across the magnetic field, meaning that it is
$\kappa_{\perp}$ that determines the $t_{TH}$ in the above
expression.  Since $\kappa_{\perp} \ll \kappa_{\parallel}$, the
thermal conduction will be inefficient in getting heat out of the
plasmoids, hence their plasma temperature will grow in time.  When
comparing the cases with and without thermal conduction, we find that
the significance of this process is in lowering the plasma temperature
(and heat content) at the locations of the $X-$points, while increasing
the temperature (and heat content) of the adjacent plasmoids (or the
$O-$point locations).  The drop in temperature at the locations of the
$X-$points means higher value of the magnetic diffusion coefficient
($\sim T^{-3/2}$) there, hence: (i) more efficient conversion of magnetic
energy into heat and kinetic energy; and, (ii) enhanced reconnection
rate, as in the case M4.  We find that in the temperature-dependent
magnetic diffusivity cases, the plasma temperature increases from
0.1 (initially) to around 0.25 inside the current sheet region, which
leads to a decrease in the magnetic diffusivity by a factor of 4 from
its initial value of $1.89 \times 10^{-5}$ down to $4.8 \times 10^{-6}$.
As the temperature rises at the locations of the $X-$points, however,
the effects of thermal conduction increase, leading to a drop in plasma
temperature on a characteristic time-scale of $t_{TH}$, which is
shorter than the characteristic time of magnetic diffusion, $t_{MD} \sim
{d_X}^2 / \eta$ (where $d_X$ is the characteristic size of the
diffusion region).  As a consequence, the drop in temperature at the
$X-$points results in an enhanced $\eta$ therein which, in turn, leads
to a decrease in $t_{MD} $.  This negative-feedback-loop proceeds
until $t_{MD} $ becomes comparable with $t_{TH}$, which is achieved
at enhanced values of $\eta$.  This ultimately leads to an enhanced
reconnection rate at the $X-$points due to the presence of thermal
conduction.  This is also the reason why the thermal conduction is
more effective in the case M3 with temperature-dependent resistivity
than in the case M1 with uniform resistivity.  

The time-dependent temperature distributions in the cases M2 and
M3 are found to be different during the later stages of the secondary
instability process.  One can see in Fig.~1(b) , Fig.~1(c), Fig.~2(a),
and Fig.~2(b) that the plasma temperature distributions inside the
current sheet for these two cases are almost identical prior to the
secondary islands appearance.  From the discussion above, the
appearance of secondary islands enhances the effects of thermal
conduction, as in the case M3, which makes the temperature
distribution along the current sheet different than that in the
case without thermal conduction (case M2).  The first and the
second plots from left to right in Fig.~5(a) and Fig.~5(b) reveal
the temperature distribution for the case M2 and the case M3,
respectively, $t = 22.534 t_A$ and $t=31.425 t_A$.  In these figures,
one can see that the hot plasma trapped in the primary plasmoid is
more spread out in the $x$-direction (at the same $y$ location) 
in the case M3 than in the case M2 at $t=31.425 t_A$.  The third
plots to the right in Fig.~5(a) and Fig.~5(b) show the temperature
distribution in the $x$-direction at $y = 4$ for the case M2 and
the case M3, respectively.  During the time period from $t=22.534
t_A$ to $t=31.425 t_A$, the highest temperature region spreads
out in the $x$-direction to a larger extent in the case M3 than in
the case M2.  This is because a thermal, front-like structure
propagates away from the center-line of the current sheet along
the x-axis in the case M3.  The enhanced heat content (and plasma
pressure) inside the plasmoid (more enhanced in the case M3 than
in the case M2 due to the heat conduction) causes a new pressure
balance to be reached at a greater width of the plasmoid in the
$x$-direction.

Note here that we have utilized anisotropic thermal conduction in
the models discussed here, and the cross-field conduction
coefficient ($\kappa_{\perp}$) is much smaller than the field-aligned
conduction coefficient ($\kappa_{\parallel}$) inside the entire current
sheet.  We find that $\kappa_{\perp}$ is comparable in value to
$\kappa_{\parallel}$ (and hence isotropic) only inside a narrow region
ranging from $x=0.4998$ to $x=0.5002$ where the magnetic field
is negligible.  The width of this region is even more narrower that
the width of the secondary current sheets that are present in our
models.  Therefore, the heat conduction is anisotropic almost
everywhere in the simulation domain, and it changes the distribution
of plasma temperature only along the magnetic field.  This is the
reason why the heat can not be conducted away in the direction
perpendicular to the current sheet.

In order to quantify the differences between the cases with (M3) and
without (M2) thermal conduction during the development of plasmoid
instabilities, we calculate the various energy contents (and fluxes)
inside the region given by: $0.4 \leq x \leq 0.6$ and $0 \leq y \leq
4$.  Due to the non-vanishing energy fluxes through the boundaries
at $x_b=0.4$ and $x_e=0.6$, the total magnetic, thermal, and kinetic
energy inside this region changes in time during the plasmoid
instability process.  The magnetic, thermal, and kinetic energy
flowing into this region through the boundaries at $x_b=0.4$ and
$x_e=0.6$ from the beginning of the simulation ($t=0$) to time $t$
is denoted as $E_{MF}(t)$, $E_{TF}(t)$, and $E_{KF}(t)$, respectively. 
(Note that these quantities may have negative signs if energy flows
out of the region.)  The magnetic, thermal, and kinetic energy
confined to this region at time $t$ is denoted as $E_{ML}(t)$, $E_{TL}(t)$,
and $E_{KL}(t)$, respectively.  The initial magnetic, thermal, and
kinetic energy at $t=0$ is denoted as $E_{MI}$, $E_{TI}$, and $E_{KI}$,
respectively.  In these notations, the dissipated magnetic energy in
the region defined by $0.4 \leq x \leq 0.6$ and $0 \leq y \leq 4$, is
given by: $E_{MD}(t) = E_{MI} + E_{MF}(t) - E_{ML}(t)$.  In the same
region, the generated thermal energy is $E_{TG}(t) = E_{TL}(t) - E_{TF}(t)
- E_{TI}$, and the generated kinetic energy is $E_{KG}(t) = E_{KL}{t}
- E_{KF}(t) - E_{KI}$.  The explicit expressions for these quantities are:

\newcommand{\ud}{\mathrm{d}} 

\begin{eqnarray}
  E_{MF}(t) = \int_{0}^{t}\!\!\!\int_{0}^{4}  [v_x(0.4,y,t) B_y(0.4,y,t)-v_y(0.4,y,t) B_x(0.4,y,t) -{}
  \nonumber\\
  \eta(0.4,y,t) (\partial_x B_y(0.4,y,t)|_{x=0.4}-\partial_y B_x(0.4,y,t))] B_y(0.4,y,t)
  \, \ud y\, \ud t - {}
  \nonumber\\
  \int_{0}^{t}\!\!\!\int_{0}^{4}  [v_x(0.6,y,t) B_y(0.6,y,t)-v_y(0.6,y,t) B_x(0.6,y,t) -{}
  \nonumber\\
  \eta(0.6,y,t) (\partial_x B_y(0.6,y,t)|_{x=0.6}-\partial_y B_x(0.6,y,t))] B_y(0.6,y,t)
  \, \ud y\, \ud t,
\end{eqnarray}
 
\begin{eqnarray}
  E_{TF}(t) =  \int_{0}^{t}\!\!\!\int_{0}^{4} (\frac {\Gamma_0 p(0.4, y,t)}{\Gamma_0-1} v_x(0.4,y,t) + F_{cond}|_{x=0.4})
  \, \ud y\, \ud t -{}
  \nonumber\\
  \int_{0}^{t}\!\!\!\int_{0}^{4} (\frac {\Gamma_0 p(0.6, y,t)}{\Gamma_0-1} v_x(0.6,y,t) + F_{cond}|_{x=0.6})
  \, \ud y\, \ud t 
\end{eqnarray}  
  
\begin{eqnarray}
  E_{KF}(t) =  \int_{0}^{t}\!\!\!\int_{0}^{4} \rho(0.4, y,t) \frac {(v_x^2(0.4,y,t)+v_y^2(0.4,y,t))}{2} v_x(0.4,y,t)
  \, \ud y\, \ud t -{}
  \nonumber\\
  \int_{0}^{t}\!\!\!\int_{0}^{4} \rho(0.6, y,t) \frac {(v_x^2(0.6,y,t)+v_y^2(0.6,y,t))}{2} v_x(0.6,y,t)
  \, \ud y\, \ud t, 
\end{eqnarray}  

\begin{equation}
  E_{ML}(t) =  \int_{0}^{4}\!\!\!\int_{0.4}^{0.6} \frac  {B_x^2(x,y,t)+B_y^2(x,y,t)}{2}
  \, \ud x\, \ud y, 
\end{equation}
 
\begin{equation}
  E_{TL}(t) =   \int_{0}^{4}\!\!\!\int_{0.4}^{0.6} \frac  {p(x,y,t)}{\Gamma_0-1}
  \, \ud x\, \ud y,   
\end{equation}
 
\begin{equation}
  E_{KL}(t) =  \int_{0}^{4}\!\!\!\int_{0.4}^{0.6} \frac  {\rho(x,y,t)(v_x^2(x,y,t)+v_y^2(x,y,t))}{2}
  \, \ud x\, \ud y     
\end{equation} 
 
The thermal energy flowing into this region consists of two parts.
The first part comes from the inward enthalpy flux through the
boundaries at $x=0.4$ and $x=0.6$.  The second part comes
from the thermal conduction effects in the $x$-direction.  The
integral calculations of the different types of energy fluxes
have been performed in IDL.  We have doubled the spatial and
temporal resolutions to check the results and make sure that
they have converged.  In Fig.~6(a) we show the time-dependent
evolution of the different kinds of energy fluxes through the
$x$-boundaries.  One can see that a substantial amount of
magnetic energy flows into the region during the period from
$t=0$ to $t=35 t_A$ during the development of the plasmoid
instability process.  On the contrary, we find that a relatively
smaller amount of thermal and kinetic energy have flown into
the region for the same time period.  Fig.~6(b) reveals that the
thermal energy conducted out of this region in the $x$-direction
due to the thermal conduction can be ignored in the case M3.
The thermal energy flux in the $x$-direction is basically brought
in by the enthalpy flux from the inflow region.  Fig.7(a), Fig.7(b),
and Fig.7(c) visualize the time-evolution of the dissipated magnetic
energy, the generated thermal energy, and the generated kinetic
energy, respectively, for the cases M2 and M3.  In these figures, one
can see that the dissipated magnetic energy is not always increasing
with time, but it decreases slightly from $t = 6.5 t_A$ to $t = 11
t_A$.  That is because the advection of magnetic flux from the inflow
region is more dominant that the dissipation of magnetic flux during
this time period, which leads to the slight increase in magnetic
energy during this time period.  We find that the generated thermal
energy increases monotonically with time during the secondary
instability process.  The generated kinetic energy is much less than
the generated thermal energy during the entire instability process,
and we find that the former is not always increasing with time.
In the cases M2 and M3 there are several peaks in the kinetic energy
from $t = 0$ to $t = 35 t_A$, with the maximum peak reached at
around $t = 3.8 t_A$ (before the secondary instabilities occur).
Therefore, there is more kinetic energy generated during the primary
tearing instability process than during the secondary instability
process.  In other words, some portion of the generated kinetic
energy has been converted to other forms of energy (thermal) after
$t = 3.8 t_A$.  The magnetic energy is found to be dissipated faster
in the case M3 than in the case M2 during the later stages of the
secondary instability process.  There is also more thermal energy
generated in the case M3 than in the case M2 during the advanced
stages of secondary instabilities.  The secondary peaks seen in the
case M3 are more pronounced (higher amplitude) than in the case M2.
Although approximately 99\% of the magnetic energy has been
eventually transformed into thermal energy, we find that the energy
conversion process during the development of the plasmoid
instabilities is rather complex.  Another key result is that there has
been more magnetic energy transformed into thermal and kinetic
energy in the case M3 than in the case M2, because the reconnection
rate is higher in the former (see discussion above). This demonstrates
once again that the thermal conduction acts to accelerates the
reconnection rate during the plasmoid instabilities process.
\textcolor{blue}{Should the simulation in the case M4 be run further
to $t=35t_A$, then one can also calculate the energy transformation
processes for this case and compare them to those in the cases M2
and M3.  Since the thermal conduction effects are stronger in the
case M4 than in the case M3, during the same time period, there
should be more magnetic energy transformed into thermal and
kinetic energy in the case M4 than in the case M3.}    

\textcolor{blue}{Note here that the purpose of this paper was to
investigate the effects of thermal conduction alone on the
reconnection dynamics, as well as on the dynamics of the secondary
instability processes.  However, should the radiative cooling be
included along with some ad-hoc volumetric heating to achieve initial
energy balance in our model, here is what we would observe over the
course of the simulation.  First, the temperature variations ($1.0-2.5
\times 10^6K$) within the current sheet in our simulation are in the
regime where the radiative loss function is weakly dependent (and
rather uniform) on the electron temperature.  Therefore, the
differences in radiative cooling will be entirely due to the electron
density (squared) fluctuations.  As can be seen in Fig.~8, the
electron density along the mid-line of the current sheet (at $x=
0.5$) has dropped almost by a factor of 2 everywhere at $t=22.5t_A$,
except for the primary magnetic island where it has increased by
$20-30\%$.  Therefore, in the former region, the radiative cooling
rate will drop by a factor of 4, whereas in the latter region it will
increase by $1.45-1.7$ times.  In other words, the primary magnetic
island will experience some enhanced radiative cooling, which will
directly compete with the time-dependent heat increase inside the
island due to the thermal conduction.  In the region where the
electron density drops almost twice compared to the initial value,
the radiative cooling rate will drop by a factor of 4, but at the same
time, the background volumetric heating (for which there is no
sensible model of how to evolve in time) will provide excessive
heating in this region.  Hence, the temperature of the secondary
current sheets, including the secondary plasmoids, will increase
with time.  Note that, although the electron density along the
mid-line of the current sheet drops with time, it is still around
$16\%$ higher at the $O$-points inside the secondary islands
than the $X$-points of the secondary current sheets, as seen in
Fig.~8.  Therefore, the radiation cooling will be around $35\%$
stronger at the $O$-points inside the secondary islands than the
$X$-points of the secondary current sheets.  Hence, due to the
excessive volumetric heating in the latter than in the former, the
maximum temperature would be observed at the reconnection
$X$-points.  In summary, the thermal conduction and the radiative
cooling will act in a competing fashion on the dynamics of the
reconnecting current sheet.  It should be noted here, however,
that the biggest uncertainty in a model that includes the radiative
cooling is in the assumed form of the volumetric heating rate. 
This heating can be constructed such that there is an energy
balance initially in the model, but then we cannot assume any
physically meaningful temporal evolution of this heating function
in time.}

\textcolor{blue}{Here, we would like to discuss briefly the relationship
between the Lundquist number, $S$, in our simulations and the obtained
rate of reconnection.  The highest $S$ in our simulations is found to
around $10^6$ as the temperature is increased to $2.5 \times 10^6 K$.
In order to ensure that the numerical resistivity is much smaller than
the physical (Spitzer-type) resistivity adopted in our models, high
numerical resolution is needed when the Lundquist number is high.
As is known, the Lundquist number in the real solar corona is around
$10^{12}$, which is beyond our computational capabilities at present.
Some recent simulation studies \cite[]{Bhattacharjee09, Huang10, Ni10,
Ni12} have indicated, however, that the reconnection rate weakly
depends on the Lundquist number as secondary instabilities appear
and $S$ exceeds a critical value.  Therefore, in some sense, $S=10^6$
can represent the physical conditions seen in reconnecting current
sheets in the solar corona.  These previous studies also demonstrate
that the reconnection rate can increase up to a high value of $\gamma
\sim 0.01$ during the secondary instability processes.  There are also
independent theoretical calculations\cite[]{Guo12}, which have shown
that the hyper-diffusivity could be an important physical process
yielding fast reconnection during the secondary instability processes.
In order to make the reconnection environment more similar to the real
solar corona, the plasma $\beta$ at the inflow boundaries in our
models is chosen to be smaller than $1.0$, and the mass density inside
the current sheet in the center is around $6$ times higher than the
inflow regions, so the plasmas in our simulations are compressible.
One of our recent works\cite[]{Ni12} has demonstrated that the
plasma $\beta$ at the inflow boundary can make the plasmoid
instability process and the reconnection rate very different, the
reconnection rate for the higher $\beta$ case is greater than the
lower $\beta$ case. The plasmas $\beta$ is high ($\beta = 6$) in the
model of Bhattachajee et al. (2009), and the plasmas densiy is uniform
and incompressible in their simulations.  Therefore, the average
magnetic reconnection rate in our simulations ($0.002 \sim 0.003$)
during the secondary instability process appears lower than the
reconnection rate measured by them. On one hand, it can be seen
in Fig.~4(a) that the heat conduction in the physical environment
similar to the solar corona will make the reconnection rate higher up
to a value that exceeds $0.003$ during the plasmoid instability
process.  On the other hand, according to the observational evidence
\cite[]{Nagashima06, Isobe09}, the estimated reconnection rate is
in the range $0.001-0.07$.  Therefore, the global reconnection rate
measured from our simulations should be fast enough to explain the
solar flares observed in the solar corona. } 

\section{CONCLUSIONS}
\label{sec:conclusions}


In this paper, we have investigated the physical effects
of temperature-dependent magnetic diffusivity and
anisotropic thermal conduction on the dynamics of plasmoid
instabilities in reconnecting current sheets in the environment
of the solar corona.
\textcolor{blue}{For the reasons presented above, we have excluded
the radiative cooling and ad-hoc volumetric heating from our
simulations.}  We have conducted five numerical experiments in 2-D
MHD systems, as presented in Table~1. The main conclusions of this
work are summarized below.
First, we have found that the plasma temperature in the current
sheet region increases with time and it becomes greater than that
in the inflow region.  Second, as secondary magnetic islands appear,
the highest temperature is not always found at the reconnection
$X$-points, but also inside the secondary islands.  One of the effects
of anisotropic thermal conduction is to decrease the temperature
of the reconnecting $X-$points and transfer the heat into the
$O-$points, the plasmoids, where it gets trapped.  Third, in the
cases with temperature-dependent magnetic diffusivity, $\eta \sim
T^{-3/2}$, the decrease in plasma temperature at the $X-$points
leads to: (i) increase in the magnetic diffusivity until the characteristic
time for magnetic diffusion becomes comparable to that of thermal
conduction; (ii) increase in the reconnection rate; and, (iii) more
efficient conversion of magnetic energy into thermal energy and
kinetic energy of bulk motions.  These results provide further
explanation of the rapid release of magnetic energy into heat
and kinetic energy seen during flares and CMEs.  We conclude that
the consideration of anisotropic thermal conduction and Spitzer-type,
temperature-dependent magnetic diffusivity, as in the real solar
corona, are crucially important for explaining the occurrence of fast
reconnection during solar eruptions.

We have also investigated the energy budget of the reconnecting
current sheets during the primary and the secondary instability
processes.  We have found that the magnetic energy is converted more
efficiently into thermal and kinetic energy during the later stages of
secondary instability process, and that the conversion process is
rather complex then.

In the future, we would like to implement open boundary conditions
in the $y$-direction along the current sheet, and to introduce a guide
field in the third dimension.  The former is necessary in order to let
the heat flow leave through boundary without being reflected back in.
The inclusion of guide magnetic field is also important, because it
will enable us to investigate how the heat trapped inside the
plasmoids ($O$-points) is carried away by the thermal conduction
in the direction of the guide field.  This type of study was not
possible in the current version of the models. 

 \acknowledgments
 This research is supported by the key Laboratory of Solar Activity
 grant KLSA2011-09, the Applied Basic Research of Yunnan Province
 in China grant 2011FB113, the NSFC grant 11147131, and the CAS
 grant KJCX2-EW-T07 at the YNAO.  I.~R. would like to acknowledge
 the financial support received from the CAS grant 2011T2J01 at the
 YNAO.  This work utilizes the NIRVANA code v3.5 developed by U.~Z.
 at the Leibniz-Institut f\"ur Astrophysik Potsdam.  The numerical
 simulations in this work were carried out at the HPC Center at the
 Kunming Institute of Botany, CAS.


\clearpage


\clearpage
\begin{table}
  \caption{Summary of Models with Used Normalization Parameters
    and Initial Equilibrium Conditions }

  \label{Parameter}
  \begin{tabular}{lcccccccc r@{.}l c}
    \hline
                &    $T_N$      &   $L_N    $     &   $B_N$         &                      $\rho_N$                     &                                     $S$                            &      Heat Conduction   \\
    Model  &$(10^7 \, K)$&$(10^7 \, m)$& $(0.01 \, T)$& $(9.576 \times 10^{-10} \, kg/m^3)$                                                                                                                \\
    \hline
    M0       &                    &                    &                      &                                                           & $\frac{4}{6}\times10^7\times T_0^{3/2}$      &                 NO  \\  
    M1       &          1        &          1        &         1           &                            1                             & $\frac{4}{6}\times10^7\times T_0^{3/2}$      &                YES  \\                     
    M2       &                    &                    &                      &                                                           & $\frac{4}{6}\times10^7\times T^{3/2}    $      &                NO  \\ 
    M3       &          1        &          1        &         1           &                            1                             & $\frac{4}{6}\times10^7\times T^{3/2}    $      &                YES  \\ 
    M4       &          1        &          1        &       0.4          &                          0.16                          & $\frac{4}{6}\times10^7\times T^{3/2}   $       &                YES  \\ 
    \hline
  \end{tabular}
\end{table}

\clearpage
\begin{figure}
  \centerline{ \includegraphics[width=0.6\textwidth, clip=]{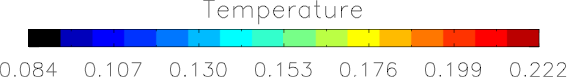} }
  \centerline{ \includegraphics[width=0.8\textwidth, clip=]{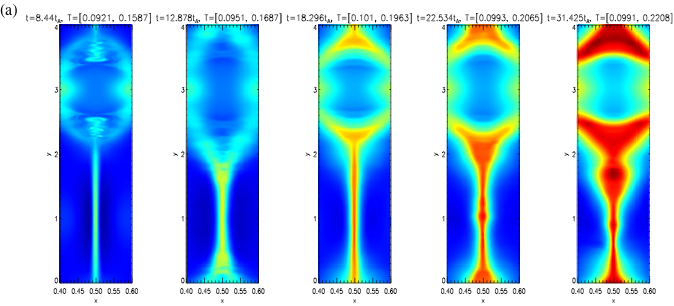} }
  \centerline{ \includegraphics[width=0.8\textwidth, clip=]{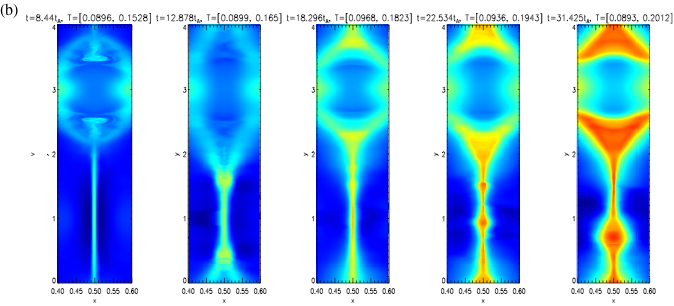} }
  \centerline {\includegraphics[width=0.8\textwidth, clip=]{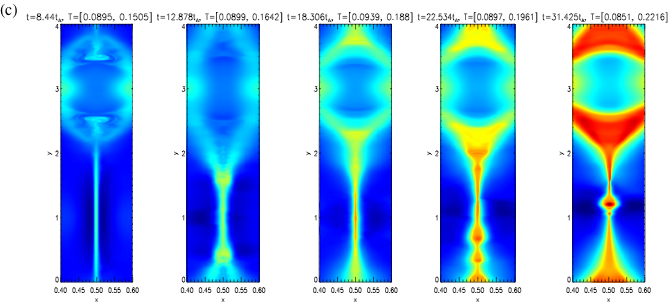} }
  \caption{The spatial distribution of plasma temperature in the case
    M0 (a), case M2 (b), and case M3 (c) at different time instants.}
  \label{fig.1}
\end{figure}

\begin{figure*}
  \centerline{\includegraphics[width=0.4\textwidth, clip=]{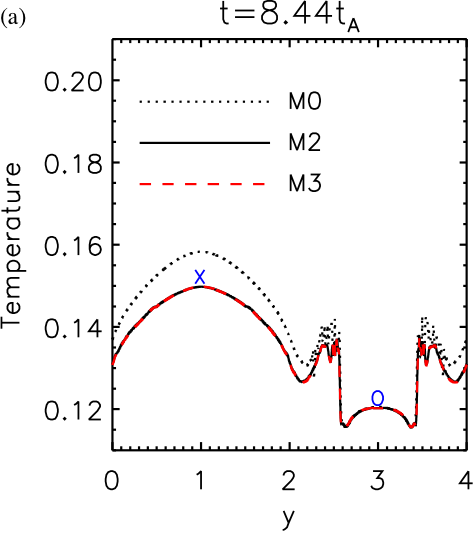}} 
  \centerline{\includegraphics[width=0.4\textwidth, clip=]{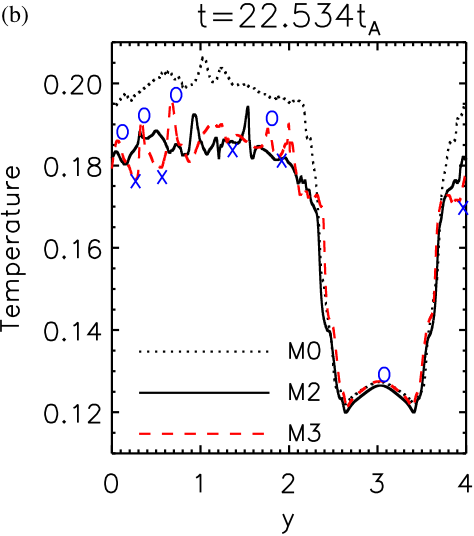}} 
  \caption{The distribution of the plasma temperature along the current
    sheet at $x=0.5$.  The dotted black line is for case M0, the solid black line corresponds to the case M2, and the dashed red line is for the case M3.  The blue 'O' signs
    indicate the locations of the $O$-points in case M3, whereas the blue 'X'
    signs mark the position of the reconnection $X$-points.  Panel (a)
    represents a time instant before the secondary islands appear, and
    panel (b) is for a time instant when the secondary islands are present.}
  \label{fig.2}
\end{figure*}

\begin{figure}
  \centerline{\includegraphics[width=0.8\textwidth, clip=]{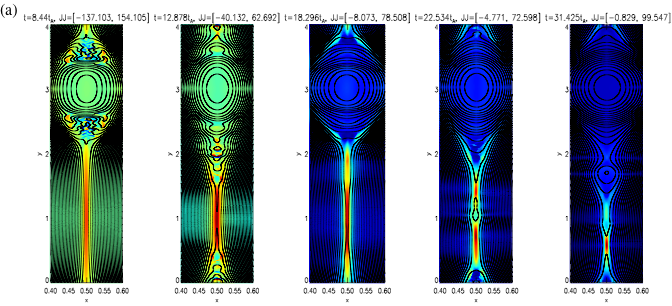}}
  \centerline{\includegraphics[width=0.8\textwidth, clip=]{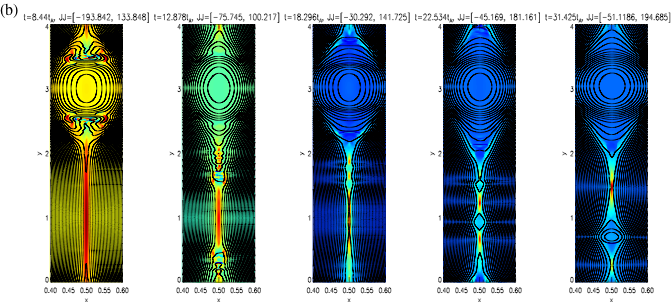}} 
  \centerline{\includegraphics[width=0.8\textwidth, clip=]{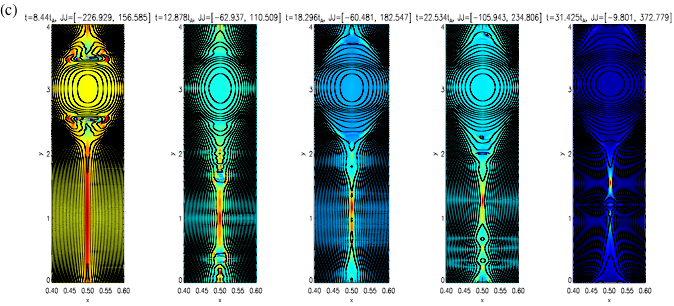}} 
  \caption{The spatial distributions of the current density and the
    magnetic flux in the case M0 (a), the case M2 (b), and the case
    M3 (c) at different time instants.}
  \label{fig.3}
\end{figure}

\begin{figure}
  \centerline{\includegraphics[width=0.5\textwidth, clip=]{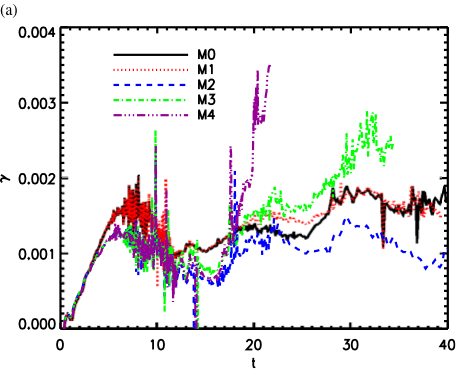}}
  \centerline{\includegraphics[width=0.5\textwidth, clip=]{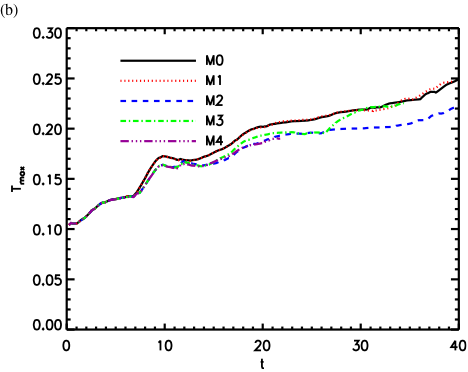}}
  \caption{(a) The time-dependent reconnection rate in all five
    cases.  (b) The time-dependent evolution of the maximum
    plasma temperature in the simulation domain in all five cases. }
  \label{fig.4}
\end{figure}

\begin{figure}
  \centerline{ \includegraphics[width=0.8\textwidth, clip=]{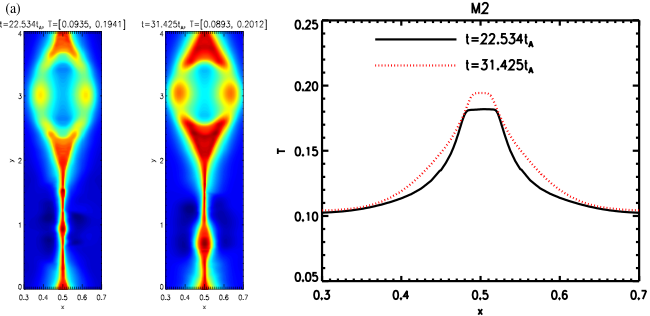} }
  \centerline{ \includegraphics[width=0.8\textwidth, clip=]{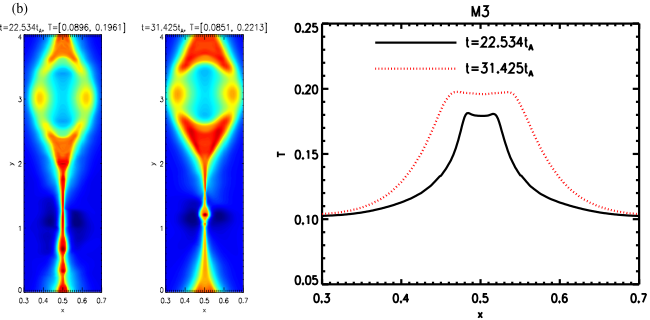} }
  \caption{The spatial distribution of the plasma temperature at time
    instants $t=22.534t_A$ and $t=31.425t_A$.  These are shown in
    the region from $x=0.3$ to $x=0.7$ for the case M2 (a) and the
    case M3 (b).  The third plot to the right in both cases illustrate
    the plasma temperature distribution at $y=4$ along the
    $x$-direction in the primary magnetic island.}
  \label{fig.5}
\end{figure}

\begin{figure}
  \centerline{\includegraphics[width=0.5\textwidth, clip=]{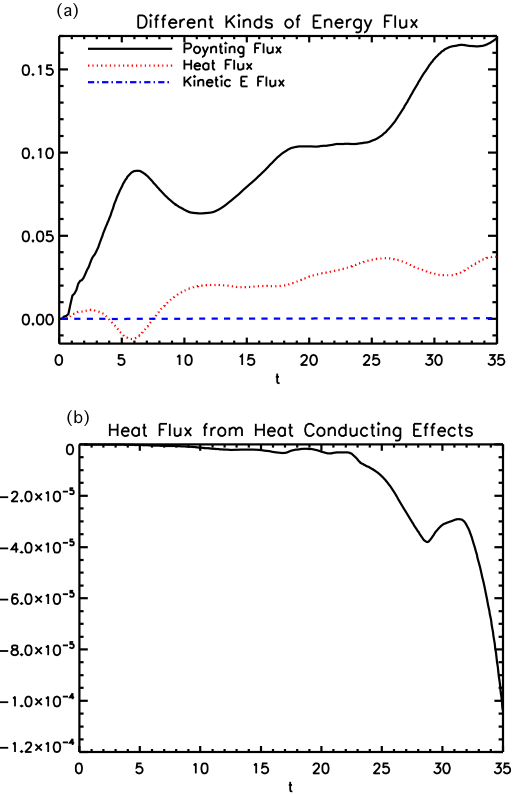}}
  \caption{(a) The time-dependent evolution of the different types
    of energy fluxes flowing into the dissipation region through the
    boundaries at $x=0.4$ and $x=0.6$ for the case M3.  (b)
    Time-dependent evolution of the heat flux conducted into
    the dissipation region for the case M3.}
  \label{fig.6}
\end{figure}

\begin{figure}
  \centerline{\includegraphics[width=0.5\textwidth, clip=]{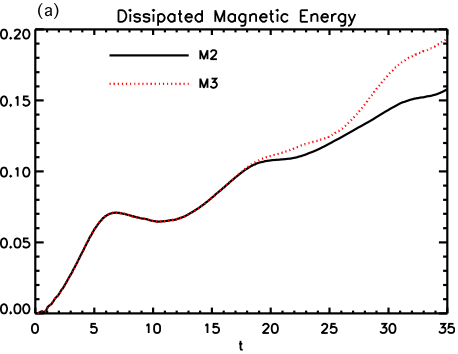} }
  \centerline{\includegraphics[width=0.5\textwidth, clip=]{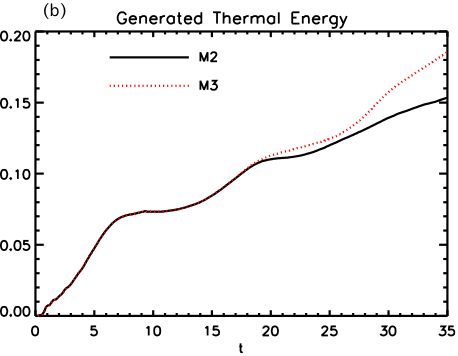}} 
  \centerline{\includegraphics[width=0.5\textwidth, clip=]{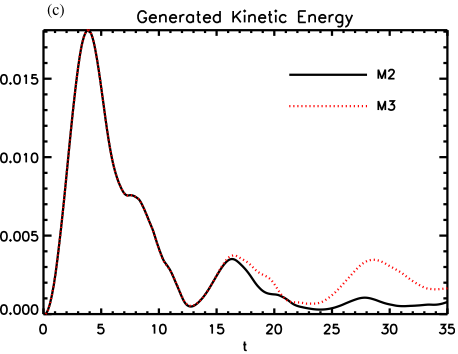} }
  \caption{The time-dependent evolution of the dissipated magnetic
    energy, the thermal energy, and the kinetic energy in the
    dissipation domain defined in the main text.}
  \label{fig.7}
\end{figure}

\begin{figure}
  \centerline{\includegraphics[width=0.4\textwidth, clip=]{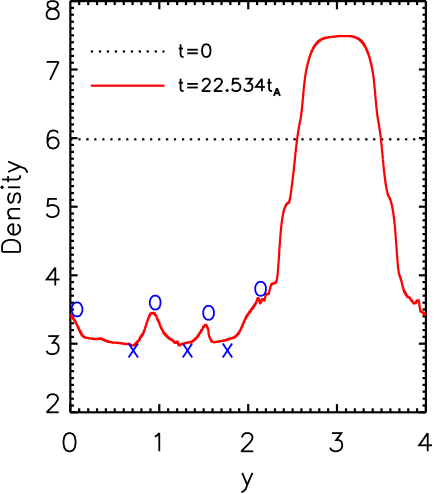}}
   \caption{The distribution of the plasma density along the current
    sheet at $x=0.5$.  The dotted black line corresponds to $t=0$,
    and the solid red line is for $t=22.534t_A$.  The blue 'O' signs
    indicate the locations of the $O$-points of the secondary
    plasmoids, whereas the blue 'X' signs mark the position of the
    reconnection $X$-points.}
  \label{fig.8}
\end{figure}
\clearpage

\end{document}